\documentclass[twocolumn,showpacs,preprintnumbers,amsmath,amssymb]{revtex4}
\usepackage{graphicx}
\usepackage{dcolumn}
\usepackage{bm}
\begin{document}
\title{Crossover Behavior in Burst Avalanches of Fiber Bundles: 
Signature of Imminent Failure}
\author{S.\ Pradhan}
\email{Pradhan.Srutarshi@phys.ntnu.no}
\author{Alex Hansen}
\email{Alex.Hansen@phys.ntnu.no}
\author{P.\ C.\ Hemmer}
\email{Per.Hemmer@phys.ntnu.no}
\affiliation{Department of Physics, Norwegian University of Science and
Technology, N--7491 Trondheim, Norway}
\begin{abstract}
Bundles of many fibers, with statistically distributed thresholds
for breakdown of individual fibers and where the load carried by a
bursting fiber is equally distributed among the surviving members,
are considered. During the breakdown process, avalanches consisting
of simultaneous rupture of several fibers occur, with a distribution
$D(\Delta)$ of the magnitude $\Delta$ of such avalanches. We show
that there is, for certain threshold distributions, a crossover behavior
of $D(\Delta)$ between two power laws $D(\Delta)\propto\Delta^{-\xi}$,
with $\xi=3/2$ or $\xi=5/2$. The latter is known to be the generic
behavior, and we give the condition for which the 
$D(\Delta)\propto\Delta^{-3/2}$ behavior is seen.  This crossover is a signal
of imminent catastrophic failure in the fiber bundle.  We find the same
crossover behavior in the fuse model.
\end{abstract}
\maketitle
A fundamental question in strength considerations of materials is when
does it fail.  Are there signals that can warn of imminent failure?  This
is of uttermost importance in e.g.\ the diamond mining industry where sudden 
failure of the mine can be extremely costly in terms of lives.  These mines
are under continuous acoustic surveillance, but at present there are no 
tell-tale acoustic signature of imminent catastrophic failure.  The same type
of question is of course also central to earth quake prediction.  We will in
this letter study signatures of imminent failure in the context of the
{\it fiber bundle model\/} \cite{Model}.  We find that if a histogram of
the number of fibers failing simultaneously is recorded over an interval
which starts sometime during the failure process, it follows a power law
with an exponent that crosses over to a very different value if the start
of the interval is close enough to the point at which the fiber bundle 
fails catastrophically. This is a clear signature of imminent failure.  We
also study the fuse model in this context and find that it behaves 
qualitatively in the same manner as the fiber bundle model.

When a weak element in a loaded material fails, the increased
stress on the remaining elements may cause further failures, and thereby
give a burst avalanche in which $\Delta$ elements fail simultaneously.
With further increase in the load new avalanches occur. A bundle of
many fibers with stochastically distributed fiber strengths, and clamped
at both ends, is a much studied model for such avalanches. In its
classical version \cite{Model}, a ruptured fiber carries no load
and the increased stresses caused by a failed element are shared equally
by all surviving fibers.

A main result \cite{HH,HH-92} for this model is that under mild restrictions
on the fiber strength distribution the expected number $D(\Delta)$
of burst avalanches of size $\Delta$ is governed by a universal power
law 
\begin{equation}
\label{powerlaw}
D(\Delta)\propto\Delta^{-\xi}
\end{equation}
for large $\Delta$, with $\xi=5/2$.

\begin{figure}
\includegraphics[width=2.4in]{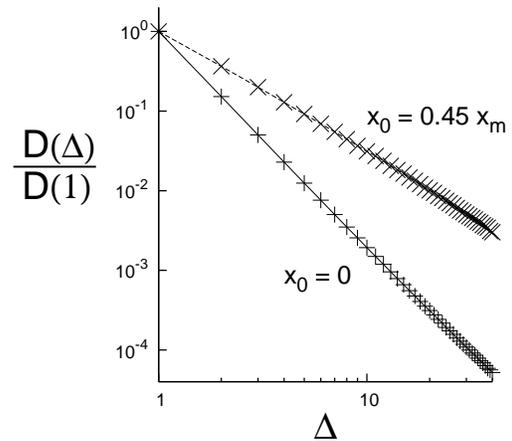}
\caption{The distribution of bursts for the strength distribution 
(\ref{uniform}) with $x_{0}=0$ and $x_{0}=0.45x_{m}$. The figure is based on 
$10000$ samples with $N=50000$ fibers.
\label{fig1}}
\end{figure}

When the load on the bundle is increased beyond a critical threshold $x_{c}$,
 the whole bundle ruptures \cite{cutoff,PH-2004}. We are interested in the 
final stages of the breakdown process, when the $N$ surviving fibers has a 
distribution of fiber strengths in the interval
($x_0$, $x_m$), with $x_0$ slightly less than  $x_{c}$).  Assume 
first that the fiber strengths in which the maximal loads  $x_{n}$ that the 
fibers  $n=1,2,\ldots N$ are able
to carry are picked independently with a probability density

\begin{eqnarray}
\lefteqn{p(x)={\textrm{Prob}}(x\,\leq\, x_{n}\,<\, x+dx)/dx}\;\nonumber\\
&\;\nonumber\\
& =(x_{m}-x_{0})^{-1} & \mbox{ for }x_{0}\leq x\leq x_{m}\;\nonumber\\
\label{uniform}
\end{eqnarray}
When $x_{0}$ is greater than the critical value $x_{c}=x_{m}/2$
the whole bundle fails fatally when the load per fiber exceeds $x_{0}$.
This can be seen as follows: The force $F(x)$ that the bundle is
able to withstand when all fibers with strengths less than $x$ have
ruptured, is proportional to the number of surviving fibers times
the strength, 
\begin{eqnarray}
\lefteqn{F(x)\propto x\;\int_{x}^{x_{m}}p(x)\; dx}\;\nonumber\\
& \nonumber\\
& =\frac{x(x_{m}-x)}{x_{m}-x_{0}} 
& \mbox{ for }x_{0}<x\leq x_{m}\;.
\label{strength}
\end{eqnarray}
When $x_{0}>x_{c}=x_{m}/2$, this is a decreasing function of $x$.
The consequence in this case is that after the first fiber ruptures
at $x=x_{0}$ the whole bundle will fail completely at once \cite{PH-2004}.
A threshold distribution in which the weakest fiber has the critical
value we call a critical threshold distribution. We want to study
the burst distribution close to this critical situation. In Fig.\ \ref{fig1}
we show results for $D(\Delta)$ in a simulation experiment on fiber
bundles with the strength distribution 
(\ref{uniform}), with $x_{0}=0.45x_{m}=0.9x_{c}$.
For comparison, results with $x_{0}=0$ are shown. In both cases $D(\Delta)$
shows a power law decay, apparently with an exponent $\xi=5/2$ for
the $x_{0}=0$ case, and an exponent $3/2$ for $x_{0}=0.45x_{m}$
(bundle with no weak fibers) \cite{PH-2004}. In this note we explain
the latter result as a crossover phenomenon.

For a bundle of many fibers the expected number of bursts of magnitude
$\Delta$ is given by \cite{HH}

\begin{figure}
\includegraphics[width=2.4in]{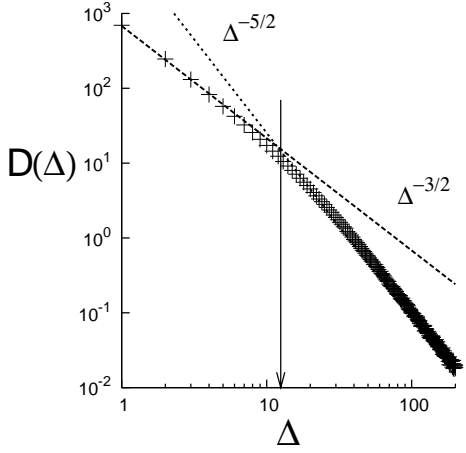}
\caption{The distribution of bursts for the strength distribution 
(Eqn.\ \ref{uniform}) with $x_{0}=0.40x_{m}$. The figure is based on 
$10000$ samples with $N=50000$ fibers. The straight lines represent two 
different power laws and the arrow locates the crossover point 
$\Delta_{c}\simeq12.5$.
\label{fig2}}
\end{figure}

\begin{eqnarray}
\lefteqn{\frac{D(\Delta)}{N}=}\nonumber\\
&\nonumber\\
& \frac{\Delta^{\Delta-1}}{\Delta!}
\int_{0}^{x_{c}}p(x)\left[1-xp(x)/Q(x)\right]
\left[xp(x)/Q(x)\right]^{\Delta-1}\nonumber\\
&\times\exp\left[-\Delta xp(x)/Q(x)\right]dx\;,\nonumber\\
\end{eqnarray}
where $Q(x)=\int_{x}^{\infty}p(y)\; dy$ is the fraction
of fibers with strength exceeding $x$. For the distribution (\ref{uniform})
this yields

\begin{eqnarray}
\lefteqn{\frac{D(\Delta)}{N}=\frac{\Delta^{\Delta-1}}{\Delta!(x_{m}-x_{0})}}
\nonumber\\
&\nonumber\\
& \times\int_{x_{0}}^{x_{c}}\frac{x_{m}-2x}{x}\left[\frac{x}{x_{m}-x}\, 
e^{-x/(x_{m}-x)}\right]^{\Delta}\; dx\;.\nonumber\\
\end{eqnarray}
Introducing the parameter
\begin{equation}
\epsilon=\frac{x_{c}-x_{0}}{x_{m}}
\end{equation}
and a new integration variable
\begin{equation}
z=\frac{x_{m}-2x}{\epsilon(x_{m}-x)}\;,
\end{equation}
we obtain 
\begin{eqnarray}
\lefteqn{\frac{D(\Delta)}{N}=\frac{2\Delta^{\Delta-1}e^{-\Delta}\,
\epsilon^{2}}{\Delta!(1+2\epsilon)}}\nonumber\\
&\nonumber\\
& \times\int_{0}^{4/(1+2\epsilon)}\frac{z}{(1-\epsilon z)(2-\epsilon z)^{2}}\; 
e^{\Delta[\epsilon z+\ln(1-\epsilon z)]}\; dz.\nonumber\\
\end{eqnarray}
For small $\epsilon$, i.e., close to the critical threshold
distribution, we expand 
\begin{equation}
\epsilon z+\ln(1-\epsilon z)
=-{\textstyle \frac{1}{2}}\epsilon^{2}z^{2}-
{\textstyle \frac{1}{3}}\epsilon^{3}z^{3}+\ldots\;,
\end{equation}
with the result

\begin{eqnarray}
\lefteqn{\frac{D(\Delta)}{N}\simeq\frac{\Delta^{\Delta-1}
e^{-\Delta}\epsilon^{2}}{2\Delta!}}\nonumber\\
&\nonumber\\
&\times\int_{0}^{4}e^{-\Delta\epsilon^{2}z^{2}/2}\; 
zdz=\frac{\Delta^{\Delta-2}e^{-\Delta}}{2\Delta!}
\left(1-e^{-8\epsilon^{2}\Delta}\right)\;.\nonumber\\
\end{eqnarray}
By use of Stirling approximation
\begin{equation}
\Delta!\simeq\Delta^{\Delta}e^{-\Delta}\sqrt{2\pi\Delta}\;,
\end{equation}
--- a reasonable approximation even for small $\Delta$ --- this may be written 
\begin{equation}
\frac{D(\Delta)}{N}\simeq(8\pi)^{-1/2}\Delta^{-5/2}
\left(1-e^{-\Delta/\Delta_{c}}\right)\;,\label{D}
\end{equation}
with 
\begin{equation}
\Delta_{c}=\frac{1}{8\epsilon^{2}}=
\frac{x_{m}^{2}}{8(x_{c}-x_{0})^{2}}\;.
\end{equation}
We see from (\ref{D}) that there is a crossover at a burst length
around $\Delta_{c}$, so that 
\begin{equation}
\frac{D(\Delta)}{N}\simeq\left\{ \begin{array}{cl}
(8/\pi)^{1/2}\epsilon^{2}\;\Delta^{-3/2} & \mbox{ for }\Delta\ll\Delta_{c}\\
(8\pi)^{-1/2}\Delta^{-5/2} & \mbox{ for }\Delta\gg\Delta_{c\;.}\end{array}
\right.
\end{equation}
For $x_{0}=0.45x_{m}$, we have $\Delta_{c}=50$, so the final
asymptotic behavior is not visible in Fig.\ \ref{fig1}. The crossover is seen
better for $x_{0}=0.40x_{m}$. In Fig.\ \ref{fig2} there is clearly a crossover
near $\Delta=\Delta_{c}=12.5$.

The phenomenon is not limited to the uniform threshold distribution.
The $\xi=3/2$ power law \cite{PH-2004} in the burst size distribution
will appear whenever a threshold distribution is non-critical, but
close to criticality. This can be seen from the expression for the
average force $F(x)\propto xQ(x)$ on the bundle. The critical value
$x_{c}$ corresponds to the maximum of $F(x)$, 
which gives $Q(x_{c})=x_{c}p(x_{c})$.
For the weakest fiber strength $x_{0}$ very close to $x_{c}$ and
$\Delta$ finite, the integrand in (4) therefore approach a constant
times $e^{-\Delta}(x_{c}-x)$. Then $D(\Delta)$ is proportional to
$\Delta^{\Delta-1}e^{-\Delta}/\Delta!\simeq(2\pi)^{-1/2}\Delta^{-3/2}$.
On the other hand, when $\Delta\gg(x_{c}-x_{0})^{-2}$ the generic
asymptotic behavior $D\propto\Delta^{-5/2}$ will follow.

\begin{figure}
\includegraphics[width=2.4in]{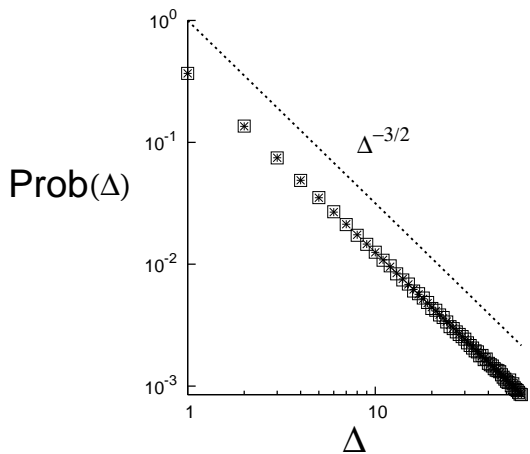}
\caption{Distribution of first burst for the critical strength distribution 
(Eqn.\ \ref{uniform}) with $x_{0}=0.50x_{m}$. The simulation results 
(squares) are based
on $10^{6}$ samples with $N=50000$ fibers. The `star' symbol stands
for the analytic result (Eqn.\ \ref{result}). 
\label{fig3}}
\end{figure}

Precisely at criticality $\Delta_{c}=\infty$  and the  $\xi=5/2$ power 
law is no 
longer present. We can demonstrate by a random walk argument that 
\emph{at criticality} the burst distribution follows a 3/2 power law. 
The load on the bundle when the $k$th fiber with strength
$x_{k}$ is about to fail is \begin{equation}
F_{k}=Q(x_{k})x_{k},\end{equation}
 and the fluctuations of this load determine the size of the bursts.
The probability $\rho(f)\, df$ that the difference $F_{k+1}-F_{k}$
is in the interval $(f,f+df)$ has been shown to be \cite{HH-92}
\begin{equation}
\rho(f)=\left\{ \begin{array}{ll}
\frac{p(x_{k})}{Q(x_{k}}\; e^{-(f+x_{k})p(x_{k})/Q(x_{k})} 
& \mbox{ for }f\geq-x_{k}\\
0 & \mbox{ for }f<-x_{ka\;.}\end{array}\right.
\end{equation}
At criticality $x\, p(x)=Q(x)$, which gives 
\begin{equation}
\rho(f)=\left\{ \begin{array}{ll}
x^{-1}\; e^{-1-f/x} & \mbox{ for }f\geq-x\\
0\mbox{ for }f<-x\;.\end{array}\right.\label{prob}
\end{equation}
This can be considered as the step probability in a random walk.
The random walk is unsymmetrical, but it is \emph{unbiased}, $<f>=0$,
as it should be at criticality. 

Using the step probability (\ref{prob}) this may be evaluated \cite{HHP}, with
the result 
\begin{equation}
\textrm{Prob}(\Delta)=\frac{\Delta^{\Delta-1}}{e^{\Delta}\;\Delta!}.
\label{result}
\end{equation}

The simulation results in Fig.\ \ref{fig3} are in excellent agreement with this
distribution. At the completion of a burst the force, i.e., the excursion
of the random walk, is larger than all previous values. Therefore
one may use this point as a new starting point to find, by the same
calculation, the distribution of the next burst, etc. Consequently
the complete burst distribution at criticality is proportional to 
(\ref{result}), i.e. essentially $\propto\Delta^{-3/2}$. 

\begin{figure}
\includegraphics[width=2.4in]{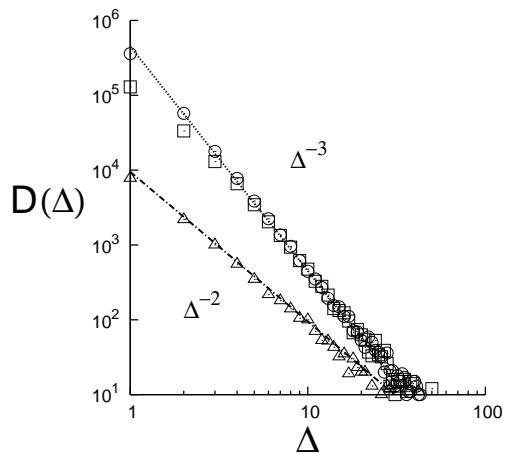}
\caption{The burst distribution in the fuse model.  System size is 
$100\times100$
and the number of samples was 300.  On the average, catastrophic failure sets
in after 2097 fuses have blown.   The circles denote the burst distribution
measured throughout the entire breakdown process.  The squares denote the
burst histrogram based on those appearing after the first 1000 fuses have blown.
The triangles denote the burst histogram after 2090 fuses have blown.  The two
straight lines indicate power laws with exponents -3.0 and -2.0 respectively.  
\label{fig4}}
\end{figure}

In order to test this crossover phenomenon in a more complex situation, we
have studied the burst distribution in the fuse model \cite{hr90}. 
The fuse model consists of a lattice where each bond is an ohmic resistor 
as long as the electrical current it carries is below a threshold value.  
If the threshold is passed, the fuse burns out irreversibly.  The threshold
$t$ of each bond is drawn from an uncorrelated distribution $p(t)$.
The lattice is placed between electrical bus bars and an increasing
current is passed through it.  Numerically, the Kirchhoff equations
are solved with a voltage difference between the bus bars set to
unity.  The ratio between current $i_j$ and threshold $t_j$ for each
bond $j$ is calculated and the bond having the largest value, $\max_j
(i_j/t_j)$ is identified and subsequently irreversibly removed.  The lattice
is a two-dimensional square one placed at 45$^\circ$ with regards to the 
bus bars.  The threshold distribution is uniform on the unit interval.  All
fuses have the same resistance.  The burst distribution follows the power
law (\ref{powerlaw}) with $\xi=3.0$, which is consistent with the value 
determined by Hansen and Hemmer \cite{hh94}.  We show the histogram in Fig.\
\ref{fig4}.  With a system size of $100\times100$, 2097 fuses blow on the
 average
before catastrophic failure sets in.  When measuring the burst distribution
only after the first 2090 fuses have blown, a different power law is found,
this time with $\xi=2.0$.  After 1000 blown fuses, on the other hand, $\xi$
remains the same as for the histogram recording the entire failure process,
see Fig.\ \ref{fig4}.

In conclusion, we have studied the burst distribution in the fiber bundle model
and shown that close to catastrophic failure it exibits a crossover behavior 
with two power law with exponents -3/2 and -5/2. In the critical situation a 
random walk argument gives a single power law with exponents -3/2. We show 
numerically, that the same crossover --- but with different values for the
exponents --- in the two-dimensional fuse model.  This crossover signals that
catastrophic failure is imminent. This signal does not hinge on observing 
rare events, and therefore has a strong potential as a useful detection tool. 

S.\ P.\ thanks the NFR for financial support through grant no.\ 166720/V30.


\end{document}